\documentclass[a4paper,aps,pra,showpacs]{revtex4}

\usepackage{amssymb}
\usepackage{graphicx}
\usepackage{amsmath}
\usepackage{amsbsy}
\usepackage{amsthm}
\usepackage{bbm}
\usepackage{bm}
\usepackage{epsfig}
\newcommand{\id}{\mathbbm{1}}

\newcommand{\be}{\begin{equation}}
\newcommand{\ee}{\end{equation}}
\newcommand{\beq}{\begin{eqnarray}}
\newcommand{\eeq}{\end{eqnarray}}
\newcommand{\bra}[1]{\ensuremath{\langle #1 |}}
\newcommand{\ket}[1]{\ensuremath{| #1 \rangle}}
\usepackage{geometry}
\begin{document}
\title{Experimentally implementable criteria revealing substructures of genuine multipartite entanglement}

\author{Marcus Huber$^1$}
\email{Marcus.Huber@univie.ac.at}
\author{Hans Schimpf$^1$}
\email{Hans.Schimpf@univie.ac.at}
\author{Andreas Gabriel$^1$}
\email{Andreas.Gabriel@univie.ac.at}
\author{Christoph Spengler$^1$}
\email{Christoph.Spengler@univie.ac.at}
\author{Dagmar Bru{\ss}$^2$}
\email{bruss@thphy.uni-duesseldorf.de}
\author{Beatrix C. Hiesmayr$^{1,3}$}
\email{Beatrix.Hiesmayr@univie.ac.at}
\affiliation{$^1$Faculty of Physics, University of Vienna, Boltzmanngasse 5, 1090 Vienna, Austria}
\affiliation{$^2$Heinrich-Heine-Universit\"at D\"usseldorf, Universit\"atsstrasse 1, D-40225 D\"usseldorf, Germany}
\affiliation{$^3$Research Center for Quantum Information, Institute of Physics, Slovak Academy of Sciences, Dubravska cesta 9, 84511 Bratislava, Slovakia}

\begin{abstract}
We present a general framework that reveals substructures of genuine multipartite entanglement. Via simple inequalities it is possible to discriminate different sets of multipartite qubit states. These inequalities are beneficial regarding experimental examinations as only local measurements are required. Furthermore, the number of observables scales favorably with system size. In exemplary cases we
demonstrate the noise resistance and discuss implementations.
\end{abstract}

\maketitle

\section{Introduction}
Quantum entanglement has been subject to intense studies in various directions for several decades and has been found in many different physical systems including systems not consisting of ordinary matter and light (see e.g. \cite{hiesmayr07,camelia}). It gave rise to several concepts for new technologies such as quantum cryptography (see e.g. \cite{cryptography}), quantum computation (e.g. \cite{qc}) or quantum teleportation (e.g. \cite{teleport}). Despite these efforts, entanglement is still far from being completely understood (for an overview, see e.g. \cite{horodecki}) and particularly in multipartite entanglement (see e.g. \cite{HHK1,HH2}) there are many unanswered questions remaining so far.

One of these open questions concerns the problem of classification of multipartite entanglement. This is usually addressed in terms of separability properties (which has been studied e.g. in Refs.~\cite{guehnewit,guehnecrit,kkbbkm}) or, in more detail, in terms of density matrix decomposition equivalence classes. The focus of this work is the latter.

While for bipartite entanglement, states can be characterized comparatively easily by means of Schmidt numbers \cite{schmidt}, multipartite entangled states have a much more complicated structure, which is only known explicitly for very special cases. For example, for three qubits, there are four distinct classes of states \cite{acin,Wstate}, namely separable ones, biseparable ones (i.e. states with two entangled particles which are separable from the third) and the two well-known classes of genuinely multipartite entangled states, the $GHZ$-state and the $W$-state (see Ref.~\cite{Wstate}). For four qubits there are nine different classes (see Ref.~\cite{vdmv}) and beyond that there is no known classification scheme.

Employing and classifying the substructure of (genuine) entanglement is of interest as current experiments are continuously improving in controlling larger and larger multipartite entangled systems, e.g. in quantum optics with photons (see e.g. Refs.~\cite{Chiuri, Wieczorek}) or with ions (see e.g. Refs.\cite{Blatt,Lavoie}) or in circuit QED of superfluid systems (see e.g. Refs.\cite{Niemczyk,Cottet}). However, it is also important to be able to determinate the entanglement class as e.g. the security of secrete sharing protocols rely on them (see Ref.~\cite{SHH3}). 

In this manuscript, we use the framework introduced in Ref.~\cite{HMGH1} in order to derive general inequalities, capable of distinguishing between different classes for multipartite qubit systems. So far, classes were usually defined with respect to SLOCC equivalence (for details see Refs.~\cite{Bastin1,Bastine1,Bastin2,Bastine2,Bastin3}), which captures whether given states can be converted into each other via stochastic local operations and classical communication. Here we pursue a different approach, defining certain equivalence classes that arise via local unitaries and permutations. This has the advantage that this class definition is Lorentz invariant (see Ref.~\cite{HFGSH1}) and we are able to develop inequalities to distinguish between these classes in an experimentally feasible way. This manuscript is organized as follows: In section \ref{sec_def} the basic terminology and definitions are introduced, so that in section \ref{sec_ineq} criteria for discriminating different classes of states can be presented, which is the main result of this work. These criteria are then tested and illustrated for a particular class of states in section \ref{sec_impl} and thoroughly discussed in section \ref{sec_dis}.

\section{Classifying multipartite entangled states\label{sec_def}}

Let us first repeat the definition for a multipartite state to be $k-$separable \cite{horodecki}. A pure $n$-partite state $\ket{\Psi}$ is called
$k$--\emph{separable} if it can be written as a product
\begin{eqnarray}
|\Psi^k\rangle=|\phi_1\rangle\otimes|\phi_2\rangle\otimes\cdots\otimes|\phi_k\rangle\
\;
\,, \quad k \leq n \,
\label{eq:ksep}
\end{eqnarray}
of $k$ states $\ket{\phi_i}$ each of which corresponds to a single
subsystem or a group of subsystems. If $k=n$ then the state is fully separable. If $k=1$, i.e. there is no such form with at
least two factors, then $\ket{\Psi}$ is $1$-separable or genuinely
$n$-partite entangled. For example, the well known GHZ state $\frac{1}{\sqrt{2}}\{|000\rangle+|111\rangle\}$ and the W-state $\frac{1}{\sqrt{3}}\{|001\rangle+|010\rangle+|100\rangle\}$ are genuinely multipartite entangled while e.g. the state $\frac{1}{\sqrt{2}}\{|00\rangle+|11\rangle\}|0\rangle$ is $2$-separable or biseparable.

For mixed states the definition of $k$-separability is straightforward: A state is $k$-separable iff the state can be decomposed into pure states \beq
\varrho&=&\sum_i\;p_i\;\ket{\Psi^k_i}\bra{\Psi^k_i} \ ,\quad p_i>0\quad\textrm{and}\quad\sum_i p_i=1 \label{deco}\eeq wherein all contained pure states are at least $k$-separable and at least one pure state is $k$-separable
   (i.e. there exists no decomposition with $(k+1)$-separable
   states and no $(k-1)$-separable state is needed in the
   decomposition). Note that a mixed state can still be partially
separable, even if the $k$ subsystems can not be split into two
groups that are not entangled with each other. E.g. the following tripartite state is biseparable ($2$-separable)
\beq
\varrho&=&\sum_i p_i \rho^i_{12}\otimes\rho^i_{3}+\sum_i q_i \rho^i_{13}\otimes\rho^i_{2}+\sum_i r_i \rho^i_{23}\otimes\rho^i_{1}\nonumber\\
&&\quad\textrm{with}\;p_i,\,q_i,\,r_i>0\quad\textrm{and}\quad\sum_{i} p_i+\sum_{j}q_j+\sum_{k}r_k=1
\eeq
where $\rho^i_{xy}$ are pure entangled bipartite states. Even though there is no bipartite splitting with respect to which the state is separable, it is considered biseparable since it can be prepared through a statistical mixture of bipartite entangled states. The nested structure of k-separability is shown in Fig. (1).
\begin{figure}[h!]
\centering 
\includegraphics[scale=0.4]{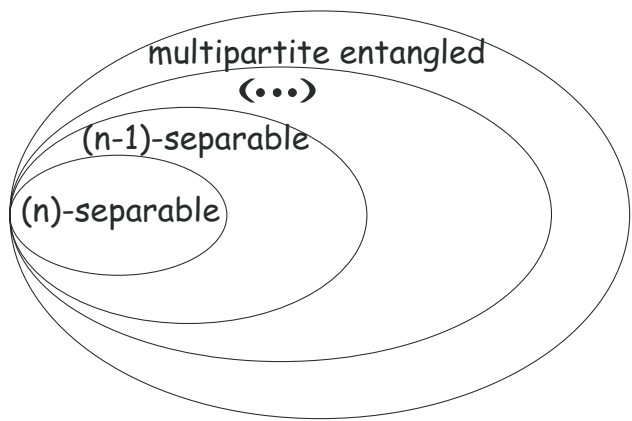}
\label{zwiebelx}
\caption{Here the nested convex structure of $k$-separability in multipartite systems is illustrated.}
\end{figure}

For pure states the question of $k$-separability can be answered by common separability criteria for bipartite systems, simply by
considering all segmentations of the $k$-partite system into two
parts. However, the same question becomes significantly more
difficult to answer for mixed states $\varrho$. In Ref.~\cite{GHH1} separability criteria for arbitrary $k$ in form of simple inequalities have been introduced.

An equivalence class (with respect to density matrix decompositions)
is defined as follows:\\
\\
\textbf{Definition 1:} A class $\mathcal{C}(\{|\Psi^k_x\rangle\})$ is defined as the set of convex mixtures of local unitary
   equivalents and permutational equivalents of $k$-separable states $|\psi_j\rangle\in\{|\Psi^k_x\rangle\}$, excluding all higher separable states ($k'>k$) within that set
\begin{equation}
\mathcal{C}(\{|\Psi^k_x\rangle\}):=\{\sum_{i,j}p_i U^i_{loc}\Pi^i|\psi_j\rangle\langle\psi_j|{\Pi^i}^\dagger{U^i_{loc}}^\dagger\}\backslash\bigcup_{i,k'>k}\mathcal{C}(\{|\Psi^{k'}_i\rangle\})\,.
\end{equation}
\\
Here, the index $x$ labels a defining property for the set $\{|\Psi_x^k\rangle\}$, e.g. the tensor rank, as in the following. The $U^i_{loc}=U^i_1\otimes U^i_2\otimes(\cdots)\otimes U_n^i$ are local unitary operators and $\Pi^i$ are permutation operators exchanging arbitrary subspaces in $|\psi_j\rangle\langle\psi_j|$.\\
In this paper we are interested in all classes of genuinely multipartite entangled states, i.e. $\mathcal{C}(\{|\Psi^1_x\rangle\})$.
It is an open question how many of such equivalence classes are needed in general for a decomposition of a given $\rho$, as in eq. (\ref{deco}). The task at hand is to identify all possible families of states $\{|\Psi^k_x\rangle\}$, which in a convex sum can build up a given density matrix $\rho$. For $n$-qubits we can introduce at least three such classes for arbitrary $n$, which can be distinguished with the mathematical framework we introduce. These three classes are defined via generalizations of the famous $GHZ$ state (Greenberger-Horne-Zeilinger, see Ref.~\cite{GHZ}), the $W$ state (see  Ref.~\cite{Wstate}) and the Dicke state with two excitations (see Ref.~\cite{Dicke}). In order to avoid confusion with the previous definition and to put them into a more general framework we label them according to their respective tensor rank (for more details on the tensor rank of multipartite states see e.g. Ref.~\cite{tensorrank}). The above definition also includes all previous classification schemes, depending on the choice of $\{|\Psi^k_x\rangle\}$, (e.g. the classification from Ref.~\cite{acin} with the two appropriate choices).\\
\noindent\textbf{Definition 2:} We define the class $\mathcal{C}(\{|\Psi^1_{(2)}\rangle\})$ of ``\textit{double states}'' for $n$
qubits (a generalization of the $GHZ$ state \cite{GHZ}) via the set 
\beq
\{\ket{\Psi^1_{(2)}}\}:=\{\ket{\psi} \in \mathbbm{C}^{2^n}:\,\ket{\psi}= \lambda_1 \bigotimes_{i=1}^n \ket{x_i} + \lambda_2 \bigotimes_{i=1}^n \ket{\overline{x}_i} \}
\eeq
where $\ket{x_i}$ are arbitrary normalized one qubit states and $\bra{\overline{x}_i}x_i\rangle=0$, such that $\ket{x_i}$ and $\ket{\overline{x}_i}$ form orthonormal bases of each single qubit system, $\lambda_1,\lambda_2\in\mathbbm{C}\setminus\{0\}$ and $|\lambda_1|^2+|\lambda_2|^2=1$.\\
\newline
\textbf{Definition 3:} We define the class $\mathcal{C}(\{|\Psi^1_{(n)}\rangle\})$ of ``\textit{$n$-tuple states}'' for
$n$ qubits (a generalization of the $W$ state \cite{Wstate}) via the set 
\beq
\{\ket{\Psi^1_{(n)}}\}:=\{\ket{\psi}\in \mathbbm{C}^{2^n}:\, \ket{\psi}= \sum_{i=1}^{n} \lambda_i \ket{W_i} \}
\eeq where
$\ket{W_i} = \bigotimes_{k\neq i}\ket{x_k}\otimes
\ket{\overline{x}_i}$ and $\lambda_i\in\mathbbm{C}\setminus\{0\}$ satisfying $\sum_{i} \left|\lambda_i\right|^2 = 1$.\\
\newline
\textbf{Definition 4:} We define the class $\mathcal{C}(\{|\Psi^1_{(n-1)}\rangle\})$  of ``\textit{$(n-1)$-tuple states}'' for
$n$ qubits (equivalent to the Dicke state for $n$ qubits \cite{Dicke} with two excitations) via the set
\begin{eqnarray}
\{|\Psi^1_{(n-1)}\rangle\}:=\{|\psi\rangle\in \mathbbm{C}^{2^n}:\,\ket{\psi}=\sqrt{\frac{2}{n(n-1)}}\sum_{i<j}|D_{ij}\rangle\}
\end{eqnarray}
where $\ket{D_{ij}} = \bigotimes_{k\neq i,j}|x_k\rangle\otimes|\overline{x}_i\rangle\otimes|\overline{x}_j\rangle$. In Ref.~\cite{tensorrank} it was shown that this
  state has tensor rank $(n-1)$.\\
Note that our definition of these classes differs from the one introduced in the context of SLOCC equivalence in
Ref.~\cite{acin}, where the classes are defined to be nested convex, such
that $W \subset GHZ$, whereas
$\mathcal{C}(\{|\Psi^1_{(n)}\rangle\})\subset
\mathcal{C}(\{|\Psi^1_{(2)}\rangle\})$ is not true. Also, in our definition there are more classes than in
Ref.~\cite{acin}, since for example the most general pure three qubit
state \beq \ket{\Psi} = \lambda_1 \ket{000} + \lambda_2 \ket{111} +
\lambda_3 \ket{001} + \lambda_4 \ket{010} + \lambda_5 \ket{100} \eeq
(see also Ref.~\cite{brun}) is contained in neither of the classes defined above when $|\lambda_i|>0\,\forall\,i$. This is discussed in more detail in section \ref{sec_dis}.\\

In the following we will show how we can distinguish our introduced classes of genuine multipartite entanglement by certain inequalities, followed by a section showing how these inequalities are experimentally implementable.

\section{Distinguishing Classes of States\label{sec_ineq}}

In the following we present our main results. All inequalities are proven in detail in the appendix (section \ref{appendix}).\\

\begin{quote}
\textbf{The ``$\mathbf{(n)}$-tuple state'' inequality I(n):} The inequality
\begin{align}\label{Wineq}\tag{I(n)}
\Re e\{\langle 0|^{\otimes n}\rho|1\rangle^{\otimes n}\}-\alpha\left(1-\langle 0|^{\otimes n}\rho|0\rangle^{\otimes n}-\langle 1|^{\otimes n}\rho|1\rangle^{\otimes n}\right) \leq 0
\end{align}
is satisfied for all biseparable states and by all states of the class $\mathcal{C}(\{|\Psi^1_{(n)}\rangle\})$, where $\alpha=\frac{3}{2}$ for $n=3$, $\alpha=1$ for $n=4$ and $\alpha=\frac{1}{2}$ for $n>4$.\\
\end{quote}
\begin{quote}
\textbf{The ``double state'' inequality I(2):} The inequality
\begin{align}\label{GHZineq}
\Re e\{\sum_{i\neq j}\langle w_i|\rho|w_j\rangle+(-1)^{n+1}\langle \overline{w}_i|\rho|\overline{w}_j\rangle\}-(n-2) \sum_i (\langle w_i|\rho|w_i\rangle+\langle \overline{w}_i|\rho|\overline{w}_i\rangle)\nonumber\\
-\sum_{i\neq j}(\langle d_{ij}|\rho|d_{ij}\rangle+\langle \overline{d}_{ij}|\rho|\overline{d}_{ij}\rangle)-\frac{n(n-1)}{2} (\langle 0|^{\otimes n}\rho|0\rangle^{\otimes n}+\langle 1|^{\otimes n}\rho|1\rangle^{\otimes n}) \leq 0\tag{I(2)}
\end{align}
is satisfied for all biseparable states and by all states of the class  $\mathcal{C}(\{|\Psi^1_{(2)}\rangle\})$, where $\ket{d_{ij}} = \ket{0}^{\otimes (i-1)}\otimes\ket{1}_i\otimes\ket{0}^{\otimes (j-i-1)}\otimes\ket{1}_j\otimes\ket{0}^{\otimes (n-j)}$, $\ket{w_{i}} = \ket{0}^{\otimes (i-1)}\otimes\ket{1}_i\otimes\ket{0}^{\otimes (n-i)}$ and an overline denotes orthonormality in all subsystems, e.g. $\ket{\overline{d}_{ij}} = \ket{1}^{\otimes (i-1)}\otimes\ket{0}_i\otimes\ket{1}^{\otimes (j-i-1)}\otimes\ket{0}_j\otimes\ket{1}^{\otimes (n-j)}$ .\\
\end{quote}
\begin{quote}
\textbf{The ``$\mathbf{(n-1)}$-tuple state'' inequality I(n-1):} The inequality
\begin{align}\label{Dickeineq}
\Re e\{\sum_{i\neq j}\langle w_i|\rho|w_j\rangle\}-(n-2) \sum_i (\langle w_i|\rho|w_i\rangle\nonumber\\
-(n-2)\sum_{i\neq j}(\langle d_{ij}|\rho|d_{ij}\rangle)-\frac{n(n-1)}{2} (\langle 0|^{\otimes n}\rho|0\rangle^{\otimes n}) \leq 0\tag{I$(n-1)$}
\end{align}
is satisfied for all biseparable states and by all states of the class  $\mathcal{C}(\{|\Psi^1_{( (n-1))}\rangle\})$.
\end{quote}
\begin{figure*}[h!]
\centering 
(a)\includegraphics[scale=0.5]{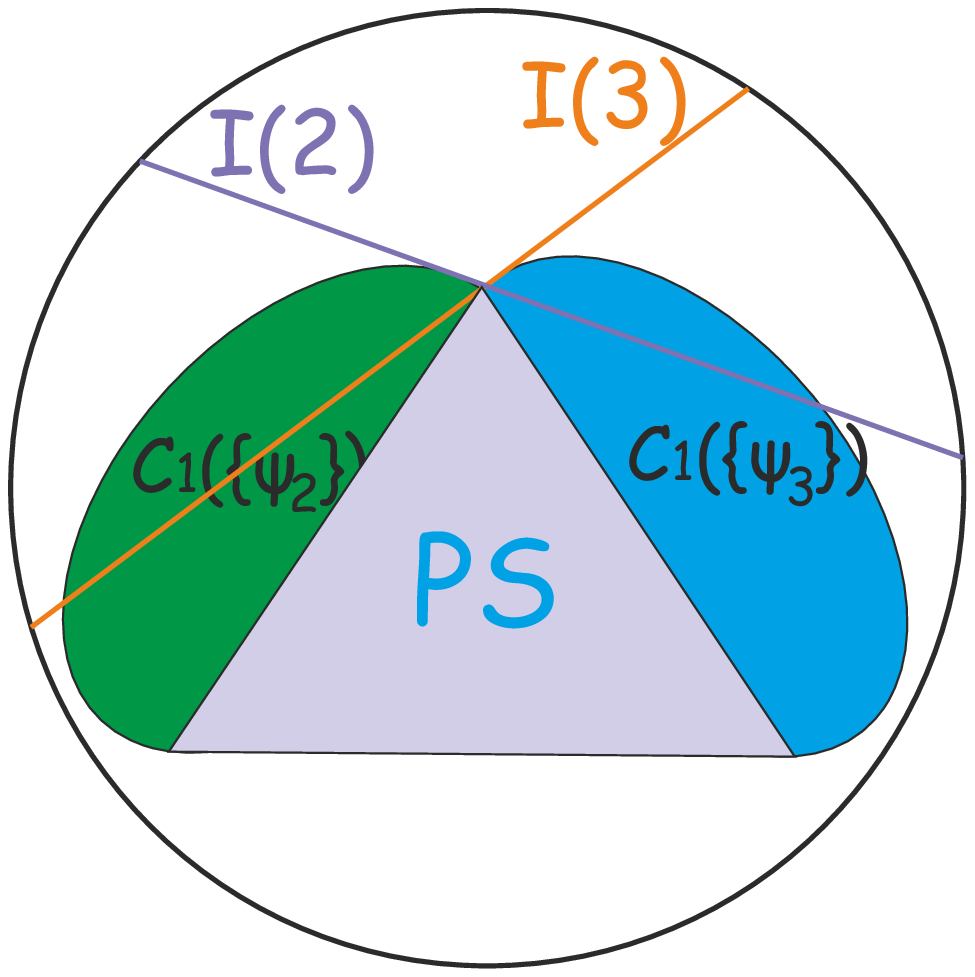}
(b)\includegraphics[scale=0.5]{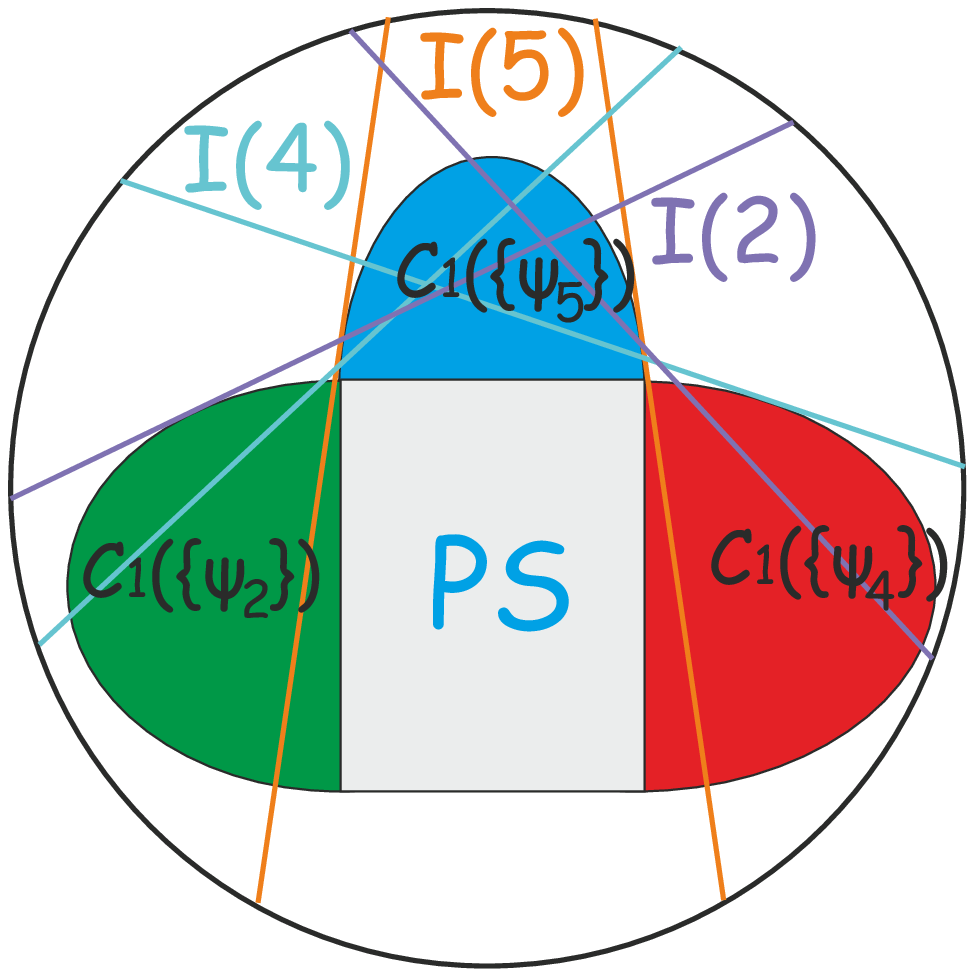}
\caption{(Color online) Here the structure of (a) three and (b) five qubit mixed states is illustrated. The three entanglement classes $\mathcal{C}(\{|\Psi^1_{(2)}\rangle\})$, $\mathcal{C}(\{|\Psi^1_{(n)}\rangle\})$ and $\mathcal{C}(\{|\Psi^1_{(n-1)}\rangle\})$ of course form convex sets together with the partially separable states, which can be excluded by our inequalities $I(2)$, $I(n)$, $I(n-1)$ and inequality $II$ from Ref.~\cite{HMGH1}. The partially separable states are subsumed as PS. The different sets are drawn completely disjoint, although it might still be possible that there is a non-vanishing overlap. Numerical analysis of tripartite systems suggests otherwise, but even if there was, it would not change any of the presented results. Also note that a two dimensional picture will always fail to incorporate all essential properties of multipartite entanglement. This illustration should facilitate the understanding of how the inequalities work.}
\label{illustr}
\end{figure*}
As these inequalities constitute only necessary conditions, their yield depends strongly on the chosen basis. So in case of unknown input states it is necessary to optimize over all local unitary operators $U = U_1\otimes U_2\otimes\cdots\otimes U_n$ to obtain optimal results. A parametrization of unitary operators which is suited very well for this task can be found in Ref.~\cite{SHH2}. In Fig. (\ref{illustr}) we illustrate the substructure of multipartite entangled states, together  with the corresponding inequalities, for three- and five-qubit states.
\section{Experimental implementation\label{sec_impl}}
\begin{figure*}[htp!]
\centering \includegraphics[scale=0.33]{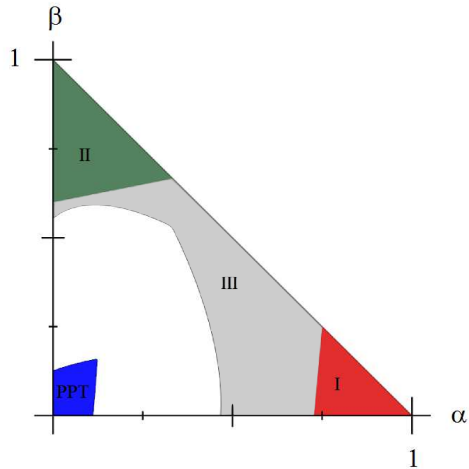}
  \caption{(Color online) Here the entanglement classes for the four qubit state $\rho=\alpha|\phi_{1}\rangle\langle\phi_{1}|+\beta|\phi_{2}\rangle\langle\phi_{2}|+\frac{1-\alpha-\beta}{16}\mathbbm{1}$ are illustrated, with $|\phi_{1}\rangle=\frac{1}{\sqrt{2}}(\ket{0000}+\ket{1111})\in\mathcal{C}\{\ket{\Psi^1_{(2)}}\}$ and $|\phi_{2}\rangle=\frac{1}{2}(\ket{1000}+\ket{0100}+\ket{0010}+\ket{0001})\in\mathcal{C}\{\ket{\Psi^1_{(4)}}\}$. For the parameter region I (red) the state is not in $\mathcal{C}(\{|\Psi^1_{(4)}\rangle\})$, for region II (green) it is not in $\mathcal{C}(\{|\Psi^1_{(2)}\rangle\})$ and for region III (grey) it is genuinely multipartite entangled, as detected by the criteria introduced in Ref.~\cite{HMGH1}. The region labeled PPT (blue) contains all states which are positive under partial transposition for the two distinct bipartitions.}
\label{exam}
\end{figure*}

It is crucial for any criteria for large systems to be implementable experimentally without having to resort to quantum state tomography, i.e. in large systems it is benefical if criteria are examinable
without knowing all entries of a density matrix. This stems from the fact that a full state tomography for $n$-qubits requires $2^{2n}$ measurement settings which for large $n$ is unfeasible. It is also very important that the criteria are locally implementable, as in large systems global measurement operations become more complex and all particles may not be available for global manipulation.\\
All three criteria to distinguish the three defined subclasses of genuine multipartite entanglement are locally implementable in experiments. This can be shown as our inequalities can be rewritten in terms of local expectation values of Pauli operators. This directly follows from the fact that our inequalities consist of a linear combination of density matrix elements, where each of which can of course be expressed in terms of local expectation values. To that end let us first introduce a compact notation in order to provide the inequalities in an elegant way:
\begin{eqnarray}
i_1i_2\cdots i_n:=\langle\sigma_{i_1}\otimes\sigma_{i_2}\otimes(\cdots)\otimes\sigma_{i_n}\rangle
\end{eqnarray}
where $\sigma_1:=\id$.\\
In this notation the elements of density matrices can be written in a compact way, e.g. for three qubits
\begin{eqnarray}
\Re e\{\langle000|\rho|111\rangle\}=xxx-yyx-yxy-xyy\,.
\end{eqnarray}
So for three qubits the ''double state''-inequality reads
\begin{eqnarray}
(xxx-yyx-yxy-xyy)-3(3-zz1-z1z-1zz)\leq 0\, ,
\end{eqnarray}
and the ''3-tuple-state''-inequality yields
\begin{eqnarray}
(1xx+xx1+x1x+1yy+y1y+yy1)\nonumber\\-\frac{9}{32}(3-zz1-z1z-1zz)-\frac{3}{16}(1-11z-1z1-z11)\leq 0\, .
\end{eqnarray}

so we see that 7 local measurement settings are required for the ''double state''-inequality and 12 for the ''triple state''- inequality as opposed to the 63 measurement settings required for a full state tomography.  In this place we notice some correspondence to the class definition via SLOCC. The witness in Ref.~\cite{GuehneBruss} uses the same observables to distinguish between the W and GHZ class (however in a different combination). The advantage of our criteria is that they generalize for systems with more parties in a straightforward way. In Fig. (\ref{exam}) we illustrate the membership to the  various entanglement classes for an example of a family of four-qubit states.
\newpage
\section{Discussion and conclusion\label{sec_dis}}
We have identified three classes of genuinely multipartite entangled states of $n$ qubits and introduced criteria which enable a simple computational and an experimental discrimination between those classes. However, as shown in Refs.~\cite{acin,vdmv}, these classes do not sufficiently characterize the set of all genuinely multipartite entangled states. In order to have a complete characterization one can use criteria to detect genuine multipartite entanglement (such as e.g. \cite{acin,HMGH1,guehnecrit,guehnewit}) to also include the complementing classes.\\
One possibility to define the complementing class is to write down the most general form of an $n$-qubit state:
\begin{equation}
|\psi_t\rangle:=\sum_{i_1,i_2,\cdots,i_n}c_{i_1,i_2,\cdots,i_n}|i_1i_2\cdots i_n\rangle
\end{equation}
All $n$-qubit states are local unitarily equivalent to this state for some choice of $c_{i_1,i_2,\cdots,i_n}$.
The equivalence class $\mathcal{C}(\{|\Psi^1_t\rangle\})$ contains all $\mathcal{C}(\{|\Psi^1\rangle\})$ for some choice of $c_{i_1,i_2,\cdots,i_n}$. So if we exclude the parameter choices for the three known classes, i.e. $\mathcal{C}(\{|\Psi^1_t\rangle\})\setminus\{\mathcal{C}(\{|\Psi^1_{(2)}\rangle\}),\mathcal{C}(\{|\Psi^1_{(n)}\rangle\}),\mathcal{C}(\{|\Psi^1_{(n-1)}\rangle\})\}$ we have the complementing set containing all other genuinely multipartite entangled states. It is completely unknown in general how many inequivalent choices for $c_{i_1,i_2,\cdots,i_n}$ are possible for $n$-partite systems, but with our proposed framework it should be possible to design an inequality for any given class.\\

To sum up we have defined three different classes of genuine multipartite entanglement for $n$ qubits, the double states (a generalization of GHZ states), the $(n)$-tuple states (a generalization of W states) and then $(n-1)$ tuple states (a generalization of certain Dicke states) which posess different physical properties and therefore provide different applications. These substructures of genuine entangled states are equivalence classes that arise via local unitaries and permutations.. We have presented three simple computable inequalities, each of which is satisfied for all biseparable states and for the class of genuine multipartite entanglement that corresponds to the inequality. Therefore, any violation of the inequality detects that the given state is not of a certain class. This certainly can be used for different applications as e.g. the success of secrete sharing protocols or quantum algorithms rely on certain class of entangled states. We have further shown that all presented criteria can be rewritten by local expectation values, thus are experimentally implementable and require far less measurement settings that full state tomography, i.e. scale favourable with the system size. Last but not least we want to stress that we presented a framework which may be generalized to any n-partite qudit system, where up to now even less is known.\\

\section{Appendix}\label{appendix}
The general concept behind all upcoming proofs is a calculation of the inequalities for the most general pure state. As all inequalities are convex the validity for mixed states is automatically given. All inequalities consist of the real parts of a certain off-diagonal density matrix elements from which diagonal elements are subtracted. In order to prove the inequalities we will first show that certain real parts of the diagonal elements cancel each other, so we can then use them freely to complete negative squares with the real parts from the off-diagonal elements and thus prove the inequality.\\
\noindent {\bf Proof of the ``double state'' inequality:}\\
The ``double state'' inequality, (\ref{GHZineq}),
\begin{eqnarray}
\sum_{i\neq j}\Re e[(\langle w_i|\rho|w_j\rangle+(-1)^{n+1}\langle \overline{w}_i|\rho|\overline{w}_j\rangle)]-\alpha\sum_i(\langle w_i|\rho|w_i\rangle+\langle \overline{w}_i|\rho|\overline{w}_i\rangle)-\nonumber\\
\beta(\langle00\cdots0|\rho|00\cdots0\rangle+\langle11\cdots1|\rho|1\cdots11\rangle)-\gamma\sum_{i\neq j}(\langle d_{ij}|\rho|d_{ij}\rangle+\langle \overline{d}_{ij}|\rho|\overline{d}_{ij}\rangle)\leq 0
\end{eqnarray}
is satisfied for all states of the form
\begin{eqnarray}
|\psi_{(2)}\rangle=\lambda_1|x_1x_2\cdots x_n\rangle+\lambda_2|y_1y_2\cdots y_n\rangle
\end{eqnarray}
where
\begin{eqnarray}
|x_i\rangle=a_i|0\rangle+\overline{a_i}|1\rangle\\
|y_i\rangle=\overline{a_i}^*|0\rangle-a_i^*|1\rangle
\end{eqnarray}
for some $\alpha,\beta,\gamma>0$\\
\noindent{\bf Proof 1:}
The individual scalar products yield
\begin{eqnarray}
\langle w_i|\psi_{(2)}\rangle=\lambda_1\prod_{n \neq i}(a_n)\overline{a}_i-\lambda_2 \prod_{n\neq i}(\overline{a^*}_n)a^*_i\\
\langle \overline{w}_i|\psi_{(2)}\rangle=\lambda_1\prod_{n \neq i}(\overline{a}_n)a_i+(-1)^{n+1}\lambda_2 \prod_{n\neq i}(a^*_n)\overline{a^*}_i\\
\langle d_{ij}|\psi_{(2)}\rangle=\lambda_1\prod_{n \neq i,j}(a_n)\overline{a}_i\overline{a}_j+\lambda_2\prod_{n\neq i,j}(\overline{a^*}_n)a^*_ia^*_j\\
\langle \overline{d}_{ij}|\psi_{(2)}\rangle=\lambda_1\prod_{n \neq i,j}(a_n)\overline{a}_i\overline{a}_j+(-1)^n\lambda_2\prod_{n\neq i,j}(\overline{a^*}_n)a^*_ia^*_j\\
\langle00\cdots0|\psi_{(2)}\rangle=\lambda_1\prod_{n}a_n+\lambda_2\prod_n\overline{a^*}_n\\
\langle11\cdots1|\psi_{(2)}\rangle=\lambda_1\prod_{n}\overline{a}_n+(-1)^n\lambda_2\prod_na^*_n\\
\end{eqnarray}
which results in
\begin{eqnarray}
\langle w_i|\psi_{(2)}\rangle\langle \psi_{(2)}|w_j\rangle=&&\underbrace{|\lambda_1|^2\prod_{n \neq i,j}(|a_n|^2)\overline{a}_ia^*_i\overline{a^*}_ja_j}_{W^0_{ij}}+\underbrace{|\lambda_2|^2 \prod_{n \neq i,j}(|\overline{a}_n|^2)\overline{a^*}_ia_i\overline{a}_ja^*_j}_{W^1_{ij}}\nonumber\\
&&-(\lambda_1\lambda^*_2 \prod_{n\neq i,j}(\overline{a}_na_n)\overline{a}^2_ia^2_j+\lambda^*_1\lambda_2 \prod_{n\neq i,j}(\overline{a^*}_na^*_n){a^*}^2_i\overline{a^*}^2_j)
\end{eqnarray}
and
\begin{eqnarray}
\langle \overline{w}_i|\psi_{(2)}\rangle\langle \psi_{(2)}|\overline{w}_j\rangle=&&\underbrace{|\lambda_1|^2\prod_{n \neq i,j}(|\overline{a}_n|^2)\overline{a^*}_ia_i\overline{a}_ja^*_j}_{\overline{W}^0_{ij}}+\underbrace{|\lambda_2|^2 \prod_{n \neq i,j}(|a_n|^2)\overline{a}_ia^*_i\overline{a^*}_ja_j}_{\overline{W}^1_{ij}}\nonumber\\
&&+(-1)^{n+1}(\lambda_1\lambda^*_2 \prod_{n\neq i,j}(\overline{a}_na_n)a^2_i\overline{a}^2_j+\lambda^*_1\lambda_2 \prod_{n\neq i,j}(\overline{a^*}_na^*_n)\overline{a^*}^2_i{a^*}^2_j)\nonumber\\
\end{eqnarray}
and
\begin{eqnarray}
|\langle w_i|\psi_{(2)}\rangle|^2=|\lambda_1|^2\prod_{n\neq i}(|a_n|^2)|\overline{a}_i|^2+|\lambda_2|^2\prod_{n\neq i}(|\overline{a}_n|^2)|a_i|^2-2\Re e[\lambda_1\lambda^*_2\prod_n(\overline{a}_na_n)]\\
|\langle \overline{w}_i|\psi_{(2)}\rangle|^2=|\lambda_1|^2\prod_{n\neq i}(|\overline{a}_n|^2)|a_i|^2+|\lambda_2|^2\prod_{n\neq i}(|a_n|^2)|\overline{a}_i|^2+(-1)^{n+1}2\Re e[\lambda_1\lambda^*_2\prod_n(\overline{a}_na_n)]
\end{eqnarray}
and
\begin{eqnarray}
|\langle00\cdots0|\psi_{(2)}\rangle|^2=\underbrace{|\lambda_1|^2\prod_{n}(|a_n|^2)}_{|M_0|^2}+\underbrace{|\lambda_2|^2\prod_n(|\overline{a}_n|^2)}_{|M_1|^2}+2\Re e[\lambda_1\lambda^*_2\prod_n(a_n\overline{a}_n)]\\
|\langle11\cdots1|\psi_{(2)}\rangle|^2=\underbrace{|\lambda_1|^2\prod_{n}(|\overline{a}_n|^2)}_{|\overline{M}_0|^2}+\underbrace{|\lambda_2|^2\prod_n(|a_n|^2)}_{|\overline{M}_1|^2}+(-1)^n2\Re e[\lambda_1\lambda^*_2\prod_n(a_n\overline{a}_n)]
\end{eqnarray}
and finally
\begin{eqnarray}
|\langle d_{ij}|\psi_{(2)}\rangle|^2=\underbrace{|\lambda_1|^2\prod_{n\neq i,j}(|a_n|^2)|\overline{a}_i|^2|\overline{a}_j|^2}_{|D^0_{ij}|^2}+\underbrace{|\lambda_2|^2\prod_{n\neq i,j}(|\overline{a}_n|^2)|a_i|^2|a_j|^2}_{|D^1_{ij}|^2}+2\Re e[\lambda_1\lambda^*_2\prod_n(\overline{a}_na_n)]\\
|\langle \overline{d}_{ij}|\psi_{(2)}\rangle|^2=\underbrace{|\lambda_1|^2\prod_{n\neq i,j}(|\overline{a}_n|^2)|a_i|^2|a_j|^2}_{|\overline{D^0}_{ij}|^2}+\underbrace{|\lambda_2|^2\prod_{n\neq i,j}(|a_n|^2)|\overline{a}_i|^2|\overline{a}_j|^2}_{|\overline{D^1}_{ij}|^2}+(-1)^n2\Re e[\lambda_1\lambda^*_2\prod_n(\overline{a}_na_n)]
\end{eqnarray}
this results for the real parts $R=2\Re e[\lambda_1\lambda^*_2\prod_n(\overline{a}_na_n)]$ in
\begin{eqnarray}
-n\alpha R+n\alpha(-1)^{n+1} R+\beta R+ (-1)^n\beta R+n(n-1)\gamma R+(-1)^nn(n-1)\gamma R=\\
=R(n\alpha((-1)^{n+1}-1)+\beta(1+(-1)^n)+n(n-1)\gamma(1+(-1)^n))
\end{eqnarray}
which is zero for odd $n$. For even $n$ and the choice $\alpha=\frac{n-2}{2}$, $\beta=\frac{n(n-2)}{4}$ and $\gamma=\frac{n-2}{4(n-1)}$ it yields
\begin{eqnarray}
=R(n\frac{n-2}{2}(-2)+\frac{n(n-2)}{4}(2)+n(n-1)\frac{n-2}{4(n-1)}(2))=0
\end{eqnarray}
Using
\begin{eqnarray}
\Re e[\langle w_i|\psi_{(2)}\rangle\langle \psi_{(2)}|w_j\rangle]-\frac{n-2}{2(n-1)}(\langle w_i|\psi_{(2)}\rangle\langle \psi_{(2)}|w_i\rangle+\langle w_j|\psi_{(2)}\rangle\langle \psi_{(2)}|w_j\rangle)\nonumber\\
=\frac{1}{n-1}\Re e[\langle w_i|\psi_{(2)}\rangle\langle \psi_{(2)}|w_j\rangle]-\underbrace{\frac{n-2}{2(n-1)}|\langle w_i|\psi_{(2)}\rangle-\langle w_j|\psi_{(2)}\rangle|^2}_{X_{ij}}
\end{eqnarray}
and
\begin{eqnarray}
\Re e[\langle\overline{w}_i|\psi_{(2)}\rangle\langle \psi_{(2)}|\overline{w}_j\rangle]-\frac{n-2}{2(n-1)}(\langle\overline{w}_i|\psi_{(2)}\rangle\langle \psi_{(2)}|\overline{w}_i\rangle+\langle\overline{w}_j|\psi_{(2)}\rangle\langle \psi_{(2)}|\overline{w}_j\rangle)\nonumber\\
=\frac{1}{n-1}\Re e[\langle\overline{w}_i|\psi_{(2)}\rangle\langle \psi_{(2)}|\overline{w}_j\rangle]-\underbrace{\frac{n-2}{2(n-1)}|\langle\overline{w}_i|\psi_{(2)}\rangle-\langle\overline{w}_j|\psi_{(2)}\rangle|^2}_{\overline{X}_{ij}}
\end{eqnarray}
the inequality now reads
\begin{eqnarray}
\sum_{i\neq j}(\frac{1}{n-1}\Re e[W^0_{ij}+W^1_{ij}+(-1)^{n+1}(\overline{W}^0_{ij}+\overline{W}^1_{ij})]-(X_{ij}+\overline{X}_{ij})-\frac{n-2}{4(n-1)}(|D^0_{ij}|^2+|D^1_{ij}|^2+|\overline{D}^0_{ij}|^2+|\overline{D}^1_{ij}|^2))\nonumber\\
-\frac{n(n-2)}{4}(|M^0|^2+|M^1|^2+|\overline{M^0}|^2+|\overline{M^1}|^2)\leq 0
\end{eqnarray}
or equivalently
\begin{eqnarray}
\sum_{i\neq j} \frac{1}{n-1}(\Re e[W^0_{ij}]-\frac{n-2}{4}|D^0_{ij}|^2-\frac{n-2}{4}|M^0|^2)\nonumber\\
\sum_{i\neq j} \frac{1}{n-1}(\Re e[W^1_{ij}]-\frac{n-2}{4}|D^1_{ij}|^2-\frac{n-2}{4}|M^1|^2)\nonumber\\
\sum_{i\neq j} \frac{1}{n-1}(\Re e[\overline{W}^0_{ij}]-\frac{n-2}{4}|\overline{D}^0_{ij}|^2-\frac{n-2}{4}|\overline{M}^0|^2)\nonumber\\
\sum_{i\neq j} \frac{1}{n-1}(\Re e[\overline{W}^1_{ij}]-\frac{n-2}{4}|\overline{D}^1_{ij}|^2-\frac{n-2}{4}|\overline{M}^1|^2)\leq 0
\end{eqnarray}
Now we can use
\begin{eqnarray}
|M^0-D^0_{ij}|^2=|M^0|^2+|D^0_{ij}|^2-2\Re e[W^0_{ij}]\\
|M^1-D^1_{ij}|^2=|M^1|^2+|D^1_{ij}|^2-2\Re e[W^1_{ij}]\\
|\overline{M}^0+(-1)^{n+1}\overline{D}^0_{ij}|^2=|\overline{M}^0|^2+|\overline{D}^0_{ij}|^2+(-1)^{n+1}2\Re e[\overline{W}^0_{ij}]\\
|\overline{M}^1+(-1)^{n+1}\overline{D}^1_{ij}|^2=|\overline{M}^1|^2+|\overline{D}^1_{ij}|^2+(-1)^{n+1}2\Re e[\overline{W}^1_{ij}]
\end{eqnarray}
which proves the inequality for $n\geq4$.\qed\\
\noindent{\bf Corollary 2}
\begin{eqnarray}
\sum_{i\neq j}\Re e[(\langle w_i|\rho|w_j\rangle+(-1)^{n+1}\langle \overline{w}_i|\rho|\overline{w}_j\rangle)]-\alpha\sum_i(\langle w_i|\rho|w_i\rangle+\langle \overline{w}_i|\rho|\overline{w}_i\rangle)-\nonumber\\
\beta(\langle00\cdots0|\rho|00\cdots0\rangle+\langle11\cdots1|\rho|1\cdots11\rangle)-\gamma\sum_{i\neq j}(\langle d_{ij}|\rho|d_{ij}\rangle+\langle \overline{d}_{ij}|\rho|\overline{d}_{ij}\rangle)\leq 0
\end{eqnarray}
is satisfied for all biseparable states for some $\alpha,\beta,\gamma>0$\\
\noindent{\bf Proof:}\\
The following inequality is satisfied by all biseparable states (as proven in Ref.\cite{HMGH1}):
\begin{eqnarray}
\sum_{i\neq j}\Re e[\langle w_i|\rho|w_j\rangle]-\sum_{i\neq j}\sqrt{\langle00\cdots0|\rho|00\cdots0\rangle\langle d_{ij}|\rho|d_{ij}\rangle}-(n-2)\sum_i\langle w_i|\rho|w_i\rangle\leq0
\end{eqnarray}
or equivalently
\begin{eqnarray}
\sum_{i\neq j}\Re e[\langle \overline{w}_i|\rho|\overline{w}_j\rangle]-\sum_{i\neq j}\sqrt{\langle11\cdots1|\rho|11\cdots1\rangle\langle \overline{d}_{ij}|\rho|\overline{d}_{ij}\rangle}-(n-2)\sum_i\langle \overline{w}_i|\rho|\overline{w}_i\rangle\leq0
\end{eqnarray}
and
\begin{eqnarray}
-\sum_{i\neq j}\Re e[\langle \overline{w}_i|\rho|\overline{w}_j\rangle]-\sum_{i\neq j}\sqrt{\langle11\cdots1|\rho|11\cdots1\rangle\langle \overline{d}_{ij}|\rho|\overline{d}_{ij}\rangle}-(n-2)\sum_i\langle \overline{w}_i|\rho|\overline{w}_i\rangle\leq0
\end{eqnarray}
using
\begin{eqnarray}
\sqrt{\langle00\cdots0|\rho|00\cdots0\rangle\langle d_{ij}|\rho|d_{ij}\rangle}\leq\frac{1}{2}(\langle00\cdots0|\rho|00\cdots0\rangle+\langle d_{ij}|\rho|d_{ij}\rangle)
\end{eqnarray}
we arrive at
\begin{eqnarray}
\sum_{i\neq j}\Re e[\langle w_i|\rho|w_j\rangle+(-1)^{n+1}\langle \overline{w}_i|\rho|\overline{w}_j\rangle]-\sum_{i\neq j}(\langle d_{ij}|\rho|d_{ij}\rangle+\langle \overline{d}_{ij}|\rho|\overline{d}_{ij}\rangle)-\nonumber\\
(n-2)\sum_i(\langle w_i|\rho|w_i\rangle+\langle \overline{w}_i|\rho|\overline{w}_i\rangle)-\frac{n(n-1)}{2}(\langle00\cdots0|\rho|00\cdots0\rangle+\langle11\cdots1|\rho|11\cdots1\rangle)\leq0
\end{eqnarray}
here we have to choose $\alpha=(n-2)$, $\beta=\frac{n(n-1)}{2}$ and $\gamma=1$.\qed\\
\noindent{\bf Corollary 3}:\\
\begin{eqnarray}
\sum_{i\neq j}\Re e[(\langle w_i|\rho|w_j\rangle+(-1)^{n+1}\langle \overline{w}_i|\rho|\overline{w}_j\rangle)]-\alpha\sum_i(\langle w_i|\rho|w_i\rangle+\langle \overline{w}_i|\rho|\overline{w}_i\rangle)-\nonumber\\
\beta(\langle00\cdots0|\rho|00\cdots0\rangle+\langle11\cdots1|\rho|1\cdots11\rangle)-\gamma\sum_{i\neq j}(\langle d_{ij}|\rho|d_{ij}\rangle+\langle \overline{d}_{ij}|\rho|\overline{d}_{ij}\rangle)\leq 0
\label{prooffinal}
\end{eqnarray}
is satisfied by both biseparable states and $\mathcal{C}(\{|\Psi^1_{(2)}\rangle\})$ for $\alpha=(n-2)$, $\beta=\frac{n(n-1)}{2}$ and $\gamma=1$.\\
\noindent{\bf Proof}:\\
It is evident that inequality (\ref{prooffinal}) is satisfied for all biseparable states as proven in Proof 2. As $\alpha$, $\beta$ and $\gamma$ are larger than what is needed for all ''double states'', we can conclude that subtracting even more positive terms does not change the validity of inequality (\ref{prooffinal}) for these states. \qed\\

\noindent {\bf Proof of the ``n-tuple state'' inequality:}
The ``n-tuple state'' inequality (\ref{Wineq})
is satisfied by all states of the form
\begin{eqnarray}
|\psi_{(n)}\rangle:=\sum_i\lambda_i|W_i\rangle
\end{eqnarray}
where
\begin{eqnarray}
|W_i\rangle:=|x_1x_2\cdots x_{i-1}y_ix_{i+1}\cdots x_n\rangle
\end{eqnarray}
and
\begin{eqnarray}
|x_i\rangle &=& a_i\;|0\rangle+\overline{a_i}\;|1\rangle\nonumber\\
|y_i\rangle &=& \overline{a_i}^*\;|0\rangle-a_i^*\;|1\rangle\;.
\end{eqnarray}
\noindent{\bf Proof 2:}
First observe that
\begin{eqnarray}
\langle 0|^{\otimes n}|\psi_{(n)}\rangle&=&\sum_i \lambda_i\; \overline{a}^*_i\; (\prod_{n\neq i} a_n)\;,\nonumber\\
\langle 1|^{\otimes n}|\psi_{(n)}\rangle&=&-\sum_i \lambda_i\; a^*_i\;(\prod_{n\neq i} \overline{a}_n)\;,
\end{eqnarray}
such that we obtain for the first term of the ``\textit{n--tuple state}'' inequality with $\rho\;=\;\sum_{i,j} \lambda_i \lambda_j^* |W_i\rangle\langle W_j|$
\begin{eqnarray}
\Re e[\langle 0|^{\otimes n}\rho|1\rangle^{\otimes n}]\;=\;-\Re e[\prod_n a_n \overline{a}^*_n+\sum_{i\neq j}\lambda_i\lambda_j^*\,{\overline{a}^{*}_i}^2 a^2_j\;(\prod_{n\neq i,j} a_n\overline{a}^*_n)]\;.
\end{eqnarray}
Let us introduce an index set $m_k=\{i_1i_2\cdots i_k\}$ and denote the complement by $m_k^C$, then we can with
\begin{eqnarray}
|d_{m_k}\rangle:=|0\rangle^{\otimes n-k}_{m_k^C}|1\rangle^{\otimes k}_{m_k}
\end{eqnarray}
rewrite the second term of the ``\textit{n--tuple state}'' inequality by
\begin{align}
1-\langle 0|^{\otimes n}\rho|0\rangle^{\otimes n}-\langle 1|^{\otimes n}\rho|1\rangle^{\otimes n}&=&\sum_{k=1}^{n-1}\sum_{m_k}\langle d_{m_k}|\rho|d_{m_k}\rangle\nonumber\\
&=&\sum_{k=1}^{n-1}\sum_{m_k}\sum_{i,j}\lambda_i\lambda_j^*\;\langle d_{m_k}|W_i\rangle\langle W_j|d_{m_k}\rangle\;,
\end{align}
i.e. we have to calculate all diagonal terms minus the first and last one. We have two different cases ($m_k/i$\dots denotes the index set $m_k$ without the index $i$)
\begin{eqnarray}
\langle d_{m_k}|W_i\rangle=\left\{\begin{array}{cl} -a^*_i\cdot\prod_{n\in m_k^C}(a_n)\cdot\prod_{l\in m_k/i}(\overline{a}_l), & \mbox{if }i\in m_k\\ \overline{a}^*_i\cdot\prod_{n\in m_k^C/i}(a_n)\cdot \prod_{l\in m_k}(\overline{a}_l),
 & \mbox{if } i\notin m_k \end{array}\right.
\end{eqnarray}
and therefore four possibilities for the products
\begin{eqnarray}
\langle W_j|d_{m_k}\rangle\langle d_{m_k}|W_i\rangle=\left\{\begin{array}{cl} \prod_{n\in m_k^C}(|a_n|^2)\prod_{l\in m_k/i,j}(|\overline{a}_l|^2)a^*_ia_j\overline{a}^*_i\overline{a}_j, & \mbox{if }i\in m_k, j\in m_k\\ -\prod_{n\in m_k^C/j}(|a_n|^2)\prod_{l\in m_k/i}(|\overline{a}_l|^2)a^*_ia_j\overline{a}^*_i\overline{a}_j, & \mbox{if } i\in m_k, j\notin m_k \\-\prod_{n\in m_k^C/i}(|a_n|^2)\prod_{l\in m_k/j}(|\overline{a}_l|^2)a^*_ia_j\overline{a}^*_i\overline{a}_j, & \mbox{if } i\notin m_k, j\in m_k\\\prod_{n\in m_k^C/\{i,j\}}(|a_n|^2)\prod_{l\in m_k}(|\overline{a}_l|^2)a^*_ia_j\overline{a}^*_i\overline{a}_j, & \mbox{if } i\notin m_k, j\notin m_k\end{array}\right.
\end{eqnarray}
We expect that certain real parts cancel each other, in detail we find that the following relation holds
\begin{eqnarray}
X^{ij}_{m_k}:=\langle W_j|d_{m_k}\rangle\langle d_{m_k}|W_i\rangle=-\langle W_j|d_{m'_{k'}}\rangle\langle d_{m'_{k'}}|W_i\rangle
\end{eqnarray}
if
\begin{eqnarray}
k'=k+1
\end{eqnarray}
and
\begin{eqnarray}
m_k\cup m'_{k+1}\setminus m_k\cap m'_{k+1}=\{i\}\;.
\end{eqnarray}
This can be proven explicitly by computing the expressions. Now we can complete all corresponding $|\langle W_j|d_{m_k}\rangle|^2+|\langle W_i|d_{m_k}\rangle|^2$ to complete negative squares.

Explicitly we observe
\begin{eqnarray}
\sum_i|\lambda_i|^2|\langle d_{\{i\}}|W_i\rangle|^2=\prod_{n}|a_n|^2\\
\sum_i|\lambda_i|^2|\langle d_{\{i\}^C}|W_i\rangle|^2=\prod_{n}|\overline{a}_n|^2
\end{eqnarray}
such that for the case $i=j$ we obtain the following negate square for the inequality under investigation
\begin{eqnarray}
&&-\sum_i |\lambda_i|^2 \Re e[\langle 0|^{\otimes n} W_i\rangle\langle W_i|1\rangle^{\otimes n}]-\alpha(\sum_i|\lambda_i|^2|\langle d_{\{i\}}|W_i\rangle|^2+\sum_i|\lambda_i|^2|\langle d_{\{i\}^C}|W_i\rangle|^2)\nonumber\\
&=&-\Re e[\prod_n a_n \overline{a}^*_n]-\alpha(\prod_{n}|a_n|^2+\prod_{n}|\overline{a}_n|^2)\nonumber\\
&=&-\frac{1}{2} |\prod_{n} a_n+\prod_{n} \overline{a}_n|^2\leq 0
\end{eqnarray}
where the last equation holds for $\alpha=\frac{1}{2}$. Thus we have proven that all terms with $i=j$ are negative.

For the terms $i\not=j$ we can proceed in a similar way
\begin{eqnarray}
&&\sum_{i,j}\left(- \Re e[\lambda_i\lambda_j^* \;\langle 0|^{\otimes n} W_i\rangle\langle W_j|1\rangle^{\otimes n}]-\alpha(|\lambda_i|^2|\langle d_{\{ij\}}|W_i\rangle|^2+|\lambda_j|^2|\langle d_{\{ij\}^C}|W_j\rangle|^2)\right)\nonumber\\
&=&\sum_{i,j}\left(-\Re e[\lambda_i\lambda_j^*\;{\overline{a}^{*}_i}^2a^2_j\,(\prod_{n\neq i,j} a_n\overline{a}^*_n)]-\alpha(|\lambda_i|^2\;\big|(-)\cdot a_i^* \overline{a}_j\; \prod_{n\not=i,j} a_n\big|^2+|\lambda_j|^2\;\big|a_i \overline{a}_j^*\; \prod_{n\not=i,j} \overline{a}_n\big|^2)\right)\nonumber\\
&=&\sum_{i,j}\left(-\frac{1}{2}\big| \lambda_i\; \overline{a}_i^* a_j(\prod_{n\not= i,j} a_n)+\lambda_j\; \overline{a}_i a_j^*(\prod_{n\not= i,j} \overline{a}_n)\big|^2\right)\;\leq 0
\end{eqnarray}
where the last equation holds for $\alpha=\frac{1}{2}$ (note that as the sum goes over all $i$ and $j$ in any sum the role of $i,j$ can be interchanged). Now we can combine both cases and the proof is complete. However, note that in the case $n=4$ the set $|d_{\{ij\}}\rangle$ and its complement are identical such that terms used to complete the negative squares for the case $i=j$ are not available for the case $i\not=j$, this we can compensate by choosing $\alpha=1$. Thus we have proven that the ``\textit{n--tuple state}'' inequality for all $n$ with $\alpha=\frac{1}{2}$ except for $n=4$ where one has to chose $\alpha=1$ holds for any state of the set $\mathcal{C}(\{|\Psi^1_{(n)}\rangle\})$.

That the inequality holds for biseparable states follows from inequality (II) presented in Ref.~\cite{HMGH1}.\qed\\
\noindent{\bf Proof of the ''$(n-1)$-tuple state'' inequality}
The inequality
\begin{eqnarray}
\sum_{i\neq j}\Re e[(\langle w_i|\rho|w_j\rangle]-(n-2)\sum_i(\langle w_i|\rho|w_i\rangle)-\nonumber\\
\frac{n(n-1)}{2}(\langle00\cdots0|\rho|00\cdots0\rangle)-(n-2)\sum_{i\neq j}(\langle d_{ij}|\rho|d_{ij}\rangle)\leq 0
\end{eqnarray}
is satisfied for all states of the form
\begin{eqnarray}
|\psi_{{ (n-1)}}\rangle=\sqrt{\frac{2}{n(n-1)}}\sum_{i<j}|D_{ij}\rangle
\end{eqnarray}
where
\begin{eqnarray}
\ket{D_{ij}} = \bigotimes_{k\neq i,j}|x_k\rangle\otimes|y_i\rangle\otimes|y_j\rangle\\
|x_i\rangle=a_i|0\rangle+\overline{a_i}|1\rangle\\
|y_i\rangle=\overline{a_i}^*|0\rangle-a_i^*|1\rangle
\end{eqnarray}
\noindent{\bf Proof 3:}
To prove the inequality we first take a look at the individual products
\begin{eqnarray}
\langle w_i|D_{xy}\rangle=\left\{\begin{array}{cl} \overline{a}_i\overline{a}_x^*\overline{a}_y^*\prod_{k \neq i,x,y} a_k, & \mbox{if }i\neq x,y\\ -a_x^*\overline{a}_y^*\prod_{k \neq x,y} a_k, & \mbox{if } i=x \\ -\overline{a}_x^*a_y^*\prod_{k \neq x,y} a_k, & \mbox{if } i=y \end{array}\right.
\end{eqnarray}
and
\begin{eqnarray}
\langle d_{ij}|D_{xy}\rangle=\left\{\begin{array}{cl} \overline{a}_i\overline{a}_j\overline{a}_x^*\overline{a}_y^*\prod_{k \neq i,j,x,y} a_k, & \mbox{if }i,j\neq x,y\\ -\overline{a}_ja_x^*\overline{a}_y^*\prod_{k \neq j,x,y} a_k, & \mbox{if }i=x, j\neq y \\ -\overline{a}_j\overline{a}_x^*a_y^*\prod_{k \neq j,x,y} a_k, & \mbox{if }i=y, j \neq x \\ -\overline{a}_ia_x^*\overline{a}_y^*\prod_{k \neq i,x,y} a_k, & \mbox{if }i\neq y, j=x\\ -\overline{a}_i\overline{a}_x^*a_y^*\prod_{k \neq i,x,y} a_k, & \mbox{if } i\neq x, j=y\\ a_x^*a_y^*\prod_{k \neq x,y} a_k, & \mbox{if }i=x,j=y \end{array}\right.
\end{eqnarray}
Now the two products
\begin{eqnarray}
|\langle\psi_{(n-1)}|w_i\rangle|^2=|\sum_{x<y} \langle w_i|D_{xy}\rangle|^2
\end{eqnarray}
and
\begin{eqnarray}
|\langle\psi_{(n-1)}|d_{ij}\rangle|^2=|\sum_{x<y} \langle d_{ij}|D_{xy}\rangle|^2
\end{eqnarray}
again have real parts which cancel each other out:
\begin{eqnarray}
\langle d_{ij}|D_{xy}\rangle\langle D_{x'y'}|d_{ij}\rangle=-\langle w_{k}|d_{mn}\rangle\langle d_{m'n'}|w_{k}\rangle
\end{eqnarray}
if
\begin{eqnarray}
\{ijijxyx'y'\}\setminus\{\{\{ij\}\cap\{xy\}\}\cup\{\{ij\}\cap\{x'y'\}\}\}\\
=\{kkmnm'n'\}\setminus\{\{\{k\}\cap\{mn\}\}\cup\{\{k\}\cap\{m'n'\}\}\}
\end{eqnarray}
from which follows that the according squared terms may be used to complete negative squares with the first part of the inequality. The condition reads
\begin{eqnarray}
\Re e[\langle w_{i}|d_{xy}\rangle\langle d_{x'y'}|w_{j}\rangle]-\frac{1}{2}(|\langle d_{ab}|D_{mn}\rangle|^2+|\langle d_{ab}|D_{m'n'}\rangle|^2)\\
=-|\langle d_{ab}|D_{mn}\rangle+\langle d_{ab}|D_{m'n'}\rangle|^2
\end{eqnarray}
when
\begin{eqnarray}
\{ijxyx'y'\}\setminus\{\{\{i\}\cap\{xy\}\}\cup\{\{j\}\cap\{x'y'\}\}\}\\
=\{abcdmnm'n'\}\setminus\{\{\{ab\}\cap\{mn\}\}\cup\{\{cd\}\cap\{m'n'\}\}\}
\end{eqnarray}
also, where the appropriate terms are not available one can use the corresponding $|\langle w_i|D_{xy}\rangle|^2$ from the inequality. By explicit, yet cumbersome, calculation one can in a straightforward way show that the inequality is indeed satisfied for all states of the class $\mathcal{C}(\{\ket{\Psi^1_{((n-1))}}\})$, analogously to the previous proofs.
As this inequality is just the linearized version of inequality (III) from Ref.~\cite{HMGH1} it also holds for all biseparable states. \qed\\

\textbf{Acknowledgements:}
We would like to thank F. Hipp for productive discussions. A. Gabriel, M. Huber and Ch. Spengler gratefully acknowledge the support of the Austrian Fund project FWF-P21947N16. D. Bru{\ss}  acknowledges
  support by DFG (Deutsche Forschungsgemeinschaft).

\end{document}